\documentclass[conference]{IEEEtran}
\IEEEoverridecommandlockouts
\usepackage{cite}
\usepackage{amsmath,amssymb,amsfonts}
\usepackage{algorithmic}
\usepackage{graphicx}
\usepackage{overpic}
\usepackage{pict2e}
\usepackage{textcomp}
\usepackage{xcolor}
\usepackage{physics}
\usepackage{hyperref} 
\hypersetup{
    colorlinks = true,
    citecolor = magenta,
    linkcolor = purple
}

\usepackage[utf8]{inputenc}
\usepackage[T1]{fontenc}

\usepackage{physics}

\def\BibTeX{{\rm B\kern-.05em{\sc i\kern-.025em b}\kern-.08em
    T\kern-.1667em\lower.7ex\hbox{E}\kern-.125emX}}

\begin{document}

\title{Promise and Challenges of Distimation\\
\thanks{This work was supported by the JST Moonshot R\&D program under Grants JPMJMS226C and JPMJMS256K, and the JSPS Bilateral Program Number JPJSBP120257718.}
}

\author{\IEEEauthorblockN{
Joshua Carlo A. Casapao\IEEEauthorrefmark{1},
Ananda G. Maity\IEEEauthorrefmark{1}\IEEEauthorrefmark{2},
Naphan Benchasattabuse\IEEEauthorrefmark{3},\\
Michal Hajdu\v{s}ek\IEEEauthorrefmark{3},
Akihito Soeda\IEEEauthorrefmark{4}\IEEEauthorrefmark{5}\IEEEauthorrefmark{6},
Rodney Van Meter\IEEEauthorrefmark{7},
and David Elkouss\IEEEauthorrefmark{1}}

\IEEEauthorblockA{\IEEEauthorrefmark{1}\textit{Networked Quantum Devices Unit, Okinawa Institute of Science and Technology Graduate University},\\
\textit{Onna-son, Okinawa 904-0495, Japan}}
\IEEEauthorblockA{\IEEEauthorrefmark{2}\textit{School of Physical Sciences, Indian Institute of Technology Goa, Ponda, Goa 403401, India}}
\IEEEauthorblockA{\IEEEauthorrefmark{3}\textit{Graduate School of Media and Governance, Keio University Shonan Fujisawa Campus, Kanagawa 252-0882, Japan}}
\IEEEauthorblockA{\IEEEauthorrefmark{4}\textit{Principles of Informatics Research Division, National Institute of Informatics, 2-1-2 Hitotsubashi, Chiyoda-ku, Tokyo, Japan}}
\IEEEauthorblockA{\IEEEauthorrefmark{5}\textit{Department of Informatics, School of Multidisciplinary Sciences, SOKENDAI}\\
\textit{(The Graduate University for Advanced Studies),
2-1-2 Hitotsubashi, Chiyoda-ku, Tokyo 101-8430, Japan}}
\IEEEauthorblockA{\IEEEauthorrefmark{6}\textit{Department of Physics, Graduate School of Science,
The University of Tokyo, 7-3-1 Hongo, Bunkyo-ku, Tokyo 113-0033, Japan}}
\IEEEauthorblockA{\IEEEauthorrefmark{7}\textit{Faculty of Environment and Information Studies, Keio University Shonan Fujisawa Campus, Kanagawa 252-0882, Japan}\\
joshuacarlo.casapao@oist.jp}
}

\maketitle

\thispagestyle{plain}
\pagestyle{plain}

\begin{abstract}
    Estimating the quality of raw entangled states and distilling high-fidelity entanglement traditionally require two separate link-layer protocols in a quantum network stack, each consuming its own share of fragile entangled pairs. 
    The recently introduced distimation concept merges these two protocols by directly extracting state estimation data from the classical syndromes generated during entanglement distillation. 
    By removing the dedicated ``test-then-distill” stage, distimation lowers the number of raw entangled pairs required to operate the network, reduces latency, and streamlines the network control plane. 
    The paradigm is especially attractive for near-term hardware platforms, where entanglement generation remains a severe bottleneck. 
    Alongside these promises, distimation introduces new engineering challenges, ranging from accurate local-device modeling to protocols for tracking time-varying sources.
    This article surveys the core principles of distimation, quantifies expected gains for realistic architectures, and outlines a research roadmap toward making distimation a standard building block in future quantum networks.
\end{abstract}


\section{Introduction}

    Scalable quantum networks, facilitated by interconnecting remote nodes with \textit{entanglement}, are the future of global communication.
    By unlocking communication capabilities beyond the limits of classical technology, this new entanglement-based paradigm also demands a shift in our conceptualization of network protocol stacks.
    Early theoretical proposals for \textit{quantum-based} data and control plane protocols~\cite{dahlberg2019link,kozlowski2020designing} have progressed to metropolitan-scale demonstrations, including entanglement distribution with platforms such as diamond nitrogen-vacancy centers and trapped‐ions~\cite{pompili2022experimental,delle2025operating}.  
    These milestones confirm the technical feasibility of end-to-end entanglement delivery while exposing the resource constraints of near-term hardware.  

    Near-term quantum communication relies on establishing a chain of short-range Bell pairs (the simplest entangled states) and using local quantum operations and classical communication (LOCC) to link the sender and receiver.
    However, Bell pair generation remains challenging due to various sources of noise and errors inherent in any physical system.

    \textit{Entanglement distillation}\footnote{In the literature, entanglement distillation is also referred to as \textit{entanglement purification}.}~\cite{bennett1996purification,fujii2009entanglement,Jansen2022enumerating}, or distillation for short, addresses inherent network imperfections by generating higher-quality Bell pairs from a larger set of lower-quality ones. 
    Because a single round is rarely sufficient, networks must determine the necessary number of rounds based on initial pair quality. 
    In the canonical ``test-then-distill'' pipeline, nodes first sacrifice freshly generated pairs to estimate noise via tomography or benchmarking~\cite{Eisert2020quantum}, running distillation only after testing to raise fidelity to application thresholds. 
    However, modest near-term hardware generation rates strictly limit Bell pair availability, making every required pair a critical resource bottleneck.

    This article reviews integrating state estimation with distillation to reduce overhead and simplify the network stack. 
    This \textit{distimation} approach (a portmanteau of distillation and estimation) is suitable for near-term platforms, where entanglement delivery is robust but limited in scale. 
    Distimation merges the two stages of the ``test-then-distill'' pipeline by harvesting noise information directly from distillation measurements, incurring zero additional quantum or classical messaging cost.
    Conceptually, this mimics classical in-band network telemetry, where real-time link information is extracted for monitoring without requiring dedicated probe packets or disrupting the underlying data plane.

    Performance-wise, distimation characterizes the Bell-diagonal entries\footnote{These denote the probability distribution of the ideal Bell state and those afflicted with Pauli errors.} of the pairs consumed for distillation.
    Although it does not provide a full tomographic reconstruction, it yields sufficient information for a link-layer service (e.g., fidelity and probability of Pauli errors).
    Distimation for Bell pairs achieves a sample complexity of $\mathcal{O}(1/\varepsilon^2)$ with respect to trace-distance error $\varepsilon$ \cite{casapao2024distimator,casapao2025distimator}, 
    making it a scalable alternative to tomography.

    Distimation immediately increases usable entanglement throughput under representative noise models and reduces link latency by eliminating testing overhead (Figure~\ref{fig:performance}, detailed later). 
    It also streamlines the connection pipeline by simultaneously handling quality assurance and error suppression (Figures~\ref{fig:stack-distimation} and~\ref{fig:high-level-schematic}).

    However, combining distillation and estimation introduces challenges: (i) extracting reliable estimates from limited syndrome statistics; (ii) remaining robust under slow hardware drift; and (iii) generalizing classical post-processing to larger protocol families. 
    Proof-of-concept protocols already demonstrate distimation for Bell-diagonal sources~\cite{casapao2024distimator,casapao2025distimator,Oka2025} and, recently, on small superconducting platforms~\cite{yokomori2025evaluation}.
    Still, an opportunity remains to formalize a systematic design methodology analogous to earlier quantum error correction work~\cite{wagner2022pauli}, where Pauli noise parameters are identified from arbitrary stabilizer code measurement statistics.

\section{Quantum network protocol stack}

    As in classical networks, a protocol stack allocates functionalities within a quantum network.
    We review this architecture to contextualize how distimation can be integrated into it.
    
    We define a \textit{link} as an abstract connection between nodes sharing an entangled resource, regardless of the underlying physical mechanism. 
    The \textit{physical layer} includes the quantum devices that facilitate these mechanisms.

\subsection{Estimation and distillation in a network stack}

    Creating long-distance entanglement relies on stitching shorter segments via entanglement swapping, but each swap reduces the resulting Bell pair's fidelity.
    This degradation makes link-level characterization essential to satisfy strict application-layer fidelity guarantees.
    Traditionally, this characterization is performed via quantum state tomography, sacrificing a subset of Bell pairs to estimate noise parameters.
    With accurate link-quality information, nodes can apply targeted solutions; for example, if raw fidelity is too low, entanglement distillation must precede swapping.
    This data also enables optimized network-wide routing, multiplexing, and scheduling, allowing nodes to plan operations that conserve quantum resources while meeting fidelity thresholds.
    Thus, reliable entanglement delivery depends on this cycle of characterization and optimization.

\subsection{Integrating distimation}

\begin{figure*}[htp]
    \centering
    \includegraphics[width=0.85\textwidth]{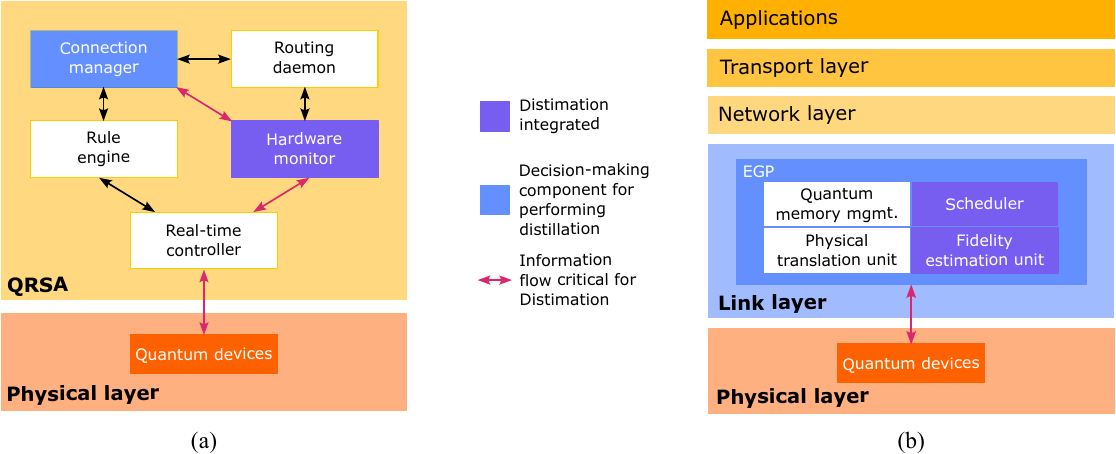}
    \caption{
    Incorporating distimation within the quantum network stacks proposed by (a) Satoh \textit{et al.} \cite{satoh-quisp} and (b) Dahlberg \textit{et al.} \cite{dahlberg2019link}. 
    (a) In the Quantum Repeater Software Architecture (QRSA) model, the responsibilities of the software are separated into five components (as discussed in the text). 
    If distimation is integrated into the hardware monitor (shown in purple), the hardware monitor plays a larger role as both a data sink and an estimation unit. 
    Information about the distillation statistics is relayed from the quantum devices (acting as the quantum network interface cards) to the hardware monitor (shown by red arrows). 
    Moreover, information about the distillation model, contained in the RuleSets generated by the connection manager, is sent to the hardware monitor.
    Periodic instructions from the hardware monitor to the real-time controller to perform a separate state characterization on the physical qubits can now be removed.
    (b) In Dahlberg \textit{et al.}'s architecture, the entanglement generation protocol (EGP) in the link layer is responsible for deciding if distillation is required. 
    Here, the distimation approach can be integrated into the EGP's fidelity estimation unit (FEU), which determines the relevant hardware parameters to meet fidelity requirements, and into its scheduler, which schedules strategies that may involve distillation. 
    Crucially, the FEU performs interspersed test rounds to ensure entanglement quality, which can be part of ``test-then-distill'' cycles.
    If distillation is necessary, these test rounds can be integrated within distillation rounds through distimation.
    }
    \label{fig:stack-distimation}
\end{figure*}

    To contextualize distimation, we consider two representative quantum network stacks (Figure~\ref{fig:stack-distimation}): the OSI-like modular stack by Dahlberg \textit{et al.}~\cite{dahlberg2019link} and the SDN-inspired RuleSet-based protocol by Satoh \textit{et al.}~\cite{satoh-quisp}.

    In the Dahlberg \textit{et al.} stack, the link layer mirrors its OSI counterpart to handle entanglement generation, distillation, and characterization.
    Its modular design simplifies integrating distimation as a link-layer service.

    In contrast, the RuleSet-based Quantum Repeater Software Architecture (QRSA) distributes these responsibilities across five software components: the connection manager, routing daemon, 
    real-time controller, rule engine, and hardware monitor. Characterization results from testing or distimation allow the hardware monitor to track fidelity and noise dynamically. 
    This data informs routing, RuleSet generation, and connection setup, eliminating dedicated test rounds.

    In both architectures, characterization becomes an integrated, potentially ``free'' service if distillation produces useful classical data, as in distimation.
    This enables real-time in-band quality monitoring of entanglement links, ensuring that long-lived entanglement chains remain viable as noise drifts over time.

    As quantum networks scale, tightly integrating state estimation with protocol execution will be essential. 
    Characterization, whether a dedicated link-layer task or part of a hybrid scheme like distimation, remains a cornerstone of reliable entanglement delivery.

\section{Distimation}

    We now introduce distimation (Figure~\ref{fig:high-level-schematic}), explain how entanglement distillation syndrome statistics can be reused for state estimation, and examine resource-accuracy trade-offs in existing distimation protocols.

\subsection{Distillation revisited}

\begin{figure*}[tp]
    \centering
    \includegraphics[width=\textwidth]{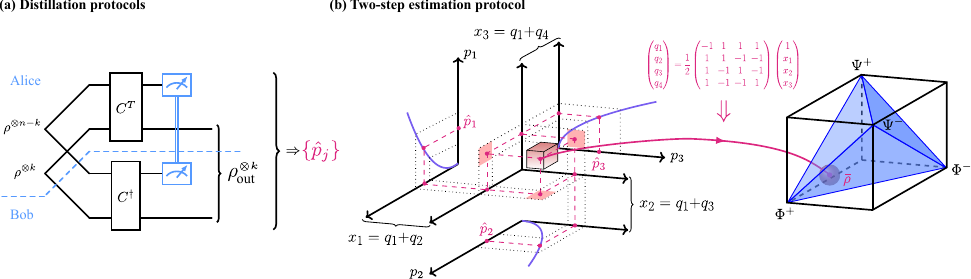}
    \caption{
    Abstract structure of a distimation protocol. 
    (a) General bilocal Clifford distillation protocols \cite{Jansen2022enumerating}. 
    Two parties share $n$ copies of an entangled state. 
    They apply a multi-qubit Clifford gate and its conjugate. 
    They then measure $n-k$ copies, share the outcomes over a classical communication channel, and keep the state depending on the coincidence pattern of the outcomes. 
    The empirical distributions of outcomes $\{\hat{p}_j\}$ exhibit strong concentration around the true underlying distribution, converging towards it at an exponential rate as the number of repetitions of the distillation protocol increases.
    These are described by the pertinent concentration inequalities relevant to distimation \cite{casapao2024distimator,casapao2025distimator}.
    (b) Schematic for distimation protocols estimating general Bell-diagonal states. 
    A two-step estimation protocol is then implemented to extract the Bell ensemble fractions from these empirical distributions. 
    In figure on the right, $\Phi^+$, $\Phi^-$, $\Psi^+$, and $\Psi^-$ are the four versions of ``pure'' (i.e., maximally entangled) Bell-diagonal states. 
    All other Bell-diagonal states fall within the convex polytope of these pure Bell-diagonal states, all corresponding to some combination of values for $q_i$'s.}
    \label{fig:low-level-general}
\end{figure*}

    In entanglement distillation, $n$ copies of low-quality entangled states (assuming sufficient generation and quantum memory) are transformed through LOCC into a smaller number of $k$ higher-fidelity states. 
    We denote this an $n$-to-$k$ \textit{protocol}.

    We consider \textit{bilocal Clifford distillation protocols} \cite{Jansen2022enumerating} (Figure~\ref{fig:low-level-general}).
    Two parties apply a Clifford gate\footnote{A gate composed of Hadamard, phase, and CNOT gates.} and its conjugate to $n$ noisy Bell pairs. 
    Each party then measures $n - k$ pairs and shares the results classically.
    Distillation is \textit{probabilistic}, post-selecting unmeasured copies based on these transmitted bits.
    These bits form the \emph{syndrome}, which correlates with underlying errors and input-state information.
    Since success depends solely on Bell-diagonal entries, only these are identifiable from syndrome statistics.

\subsection{Assumptions} 
    Distimation assumes that entanglement distillation is necessary and viable: the physical layer supplies raw pairs with fidelity insufficient for the target application, and the hardware can buffer enough raw pairs within coherence times to execute the required local gates, measurements, and classical messaging, yielding distilled states.
    
    Furthermore, we assume that dominant local noise processes (e.g., Pauli/depolarizing channels and measurement bit-flips) are well-characterized, justified because entanglement generation is qualitatively more expensive than local quantum hardware operations.
    This informs hardware-software co-design, aligning distimation algorithms with hardware constraints.

\subsection{Conceptual framework for distimation} 

\begin{figure}[t]
    \centering
    \includegraphics[width=0.475\textwidth]{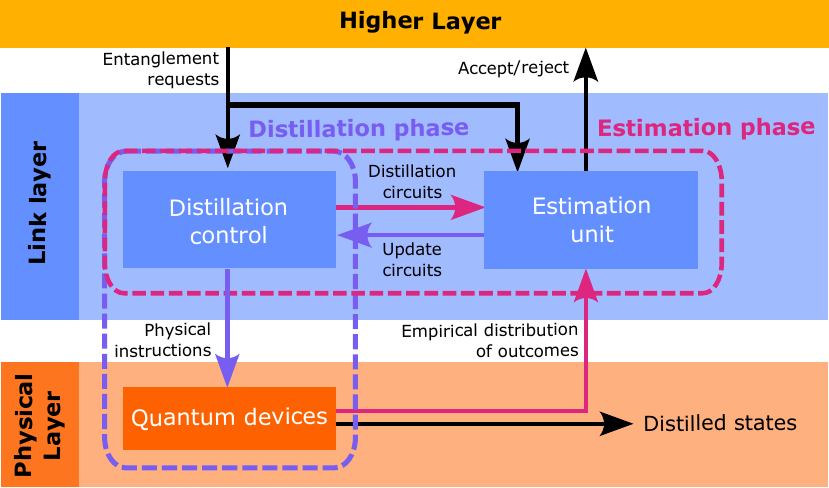}
    \caption{
    High-level schematic of the role of distimation in a quantum network. 
    Higher layers send entanglement requests to lower layers, specifying explicit quality requirements such as latency and entanglement fidelity.
    The distillation control mechanism at the link layer determines the amount of distillation necessary to ensure entanglement quality, that is, to meet the latency and fidelity requirements. 
    During the distillation phase, instructions to perform distillation are sent by the link layer to the quantum hardware platform (physical layer). 
    In the estimation phase, information about the distillation circuits, together with the empirical distribution of outcomes (i.e., the syndrome statistics) from distillation, is given as input to the estimation unit. 
    Based on the estimation, the estimation unit accepts or rejects the entanglement requests.
    The estimation results can also be used to proactively fine-tune the distillation protocol such that entanglement requests are met.
    }
    \label{fig:high-level-schematic}
\end{figure}

    As illustrated in Figure~\ref{fig:high-level-schematic}, distimation begins by receiving user entanglement requests (e.g., resource count, target fidelity, and accuracy guarantees) that support higher-layer network functionality. 
    Control mechanisms determine the necessary distillation protocol to meet the request. 
    Nodes then execute the protocol, record corresponding syndrome statistics, and deliver successful states to the higher layer.

    Once sufficient syndrome statistics have been gathered, the estimation phase starts.
    Because nodes already possess these local syndrome statistics from the distillation step, distimation requires no additional classical messaging to exchange measurement outcomes---an overhead reduction compared to traditional tomography.
    The required sample size can be specified as metadata during initial link scheduling.
    Nodes feed the circuit details and empirical syndrome distributions into an estimation unit to estimate the raw states (Figure~\ref{fig:low-level-general})~\cite{casapao2024distimator,casapao2025distimator}. 
    While current post-processing relies on direct inversion, this can be augmented in future studies with sophisticated techniques such as Bayesian analysis (see discussion of potential challenges later).
    Assuming characterized local operations, the estimates of raw states can also be propagated through the protocol to estimate the final distilled states. 
    These estimates are then fed back to the control plane to optimize future protocol selection and resource management.

\subsection{Overview of existing distimation protocols}

    We now discuss the estimation phase of several representative distimation protocols.
    Estimation goals vary by application and resource availability, ranging in difficulty from (i) estimating fidelity, to (ii) characterizing state matrix diagonals, and (iii) full-state characterization.

    A particularly simple example is the \textit{isotropic state}, which can be intuitively understood as a Bell pair with an equal probability of experiencing a Pauli $X$, $Y$, or $Z$ error. 
    This equal-error distribution allows the state to be completely characterized by a single noise parameter, $w$.
    
\begin{figure}[t]
    \centering
    \includegraphics[width=0.475\textwidth]{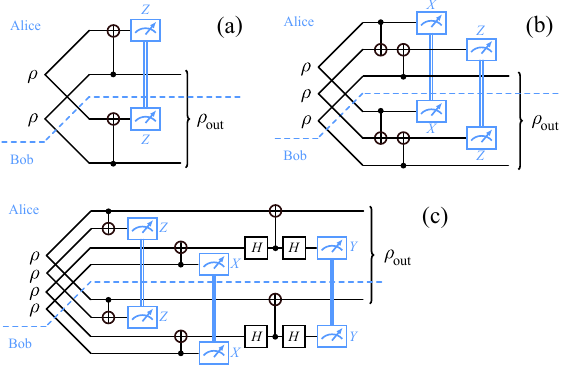}
    \caption{
    Bilocal Clifford entanglement distillation protocols studied for distimation.
    (a) The 2-to-1 BBPSSW protocol \cite{bennett1996purification}.
    (b) The 3-to-1 double selection protocol \cite{fujii2009entanglement}, with computational and Hadamard measurement settings. 
    (c) A 4-to-1 entanglement distillation protocol \cite{Oka2025}, with three distinct measurement settings.
    The blue solid lines represent the classical communication channels between the two users.
    In all protocols, a successful distillation is heralded if corresponding marginal measurement outcomes coincide.
    }
    \label{fig:dist-circuit}
\end{figure}

    The standard 2-to-1 bilocal Clifford BBPSSW protocol distills isotropic states \cite{bennett1996purification} (Figure~\ref{fig:dist-circuit}).
    Its success probability decreases monotonically with noise $w$. By repeating the protocol, parties obtain an empirical success probability they can invert to find $w$ \cite{casapao2024distimator}. 
    Finite-statistic estimation errors naturally propagate to the $w$ estimate.
    While distimation can be made robust to memory decoherence and gate errors \cite{casapao2024distimator}, analytical inversion often becomes intractable under noise, requiring numerical methods.

    Isotropic states are convenient for theoretical analysis, but experimentally generated states often deviate from this form. 
    A simple modification extends this method to characterize the diagonal coefficients---the probabilities of the ideal state and the Pauli error states---of arbitrary states in the Bell basis.
    Let $\mathbf{q}=(q_1,q_2,q_3,q_4)$ denote these coefficients. 
    Because they represent the probability of observing the corresponding Bell state if the state is measured in the Bell basis, they sum to one and equal the fidelity of each Bell state.

    The solution proposed in \cite{casapao2024distimator} uses three independent probabilistic distillation protocols. 
    Figure~\ref{fig:low-level-general} illustrates the general idea.
    Each is a variation of the low-depth BBPSSW protocol. 
    The estimation algorithm involves two steps: (i) extract an intermediate length-$3$ vector $\mathbf{x}=(x_1,x_2,x_3)$ describing the success probabilities of the three protocols; and (ii) transform this vector into an estimation of the length-$4$ diagonal vector. 
    The first step is inspired by the inversion step for isotropic states; however, the success probability for each distillation protocol is a monotonic function of a specific $x_i$ instead. 
    Each $x_i$ is the sum of $q_1$ and one of the other diagonal entries $q_{i+1}$ (i.e., $x_i=q_1+q_{i+1}$). 
    As with BBPSSW, the estimation error of $x_i$ propagates to the overall accuracy of the diagonal entries.

    We again note that the success probabilities depend only on the Bell-diagonal entries.
    Although distimation cannot characterize off-diagonal elements, it does not assume the state's form. 
    All bilocal Clifford-based distimation methods share this property.
    
    Beyond BBPSSW, Oka \textit{et al.}~\cite{Oka2025} studied distimation with a larger 4-to-1 protocol (Figure~\ref{fig:dist-circuit}), though a rigorous sample complexity analysis remains lacking.
    Alternatively, the authors in~\cite{casapao2025distimator} analyze the 3-to-1 double selection protocol~\cite{fujii2009entanglement}.
    Intuitively, using marginal and \textit{joint} measurement outcomes from this protocol enables the recovery of $\mathbf{q}$.
    While rigorous analysis is substantially more involved than in \cite{casapao2024distimator}, a conceptually similar inversion strategy applies.

\subsection{Performance evaluation}

    We compare the distimation workflow with a conventional ``test‑then‑distill'' baseline, which first reserves a block of freshly generated Bell pairs for tomography and then applies a single BBPSSW distillation round to the remainder. In contrast, distimation continuously reuses the syndrome bits produced during each distillation attempt. 
    We allocate the same number of available Bell pairs to both strategies. 
    Specifically, both operate for the same amount of time, on the order of minutes. 
    Both strategies also target the same accuracy in the noise parameter estimate. 
    We assume the latency due to local gates, measurements, and classical post‑processing to be negligible relative to entanglement generation.

    We employ isotropic states as a didactic benchmark. 
    However, experimental sources often deviate from this ideal. 
    The reported sample counts here thus underestimate the requirements of fully general multiparameter sources for both tomography and distimation. 
    Although absolute sampling overhead increases with additional parameters, the relative advantage in resource‑scarce regimes persists; see, e.g.~\cite{casapao2024distimator,casapao2025distimator} for additional comparisons.
    We make this choice to cleanly expose the structural resource trade‑off.

\begin{figure}[htp]
    \centering
    \includegraphics[width=0.5\textwidth]{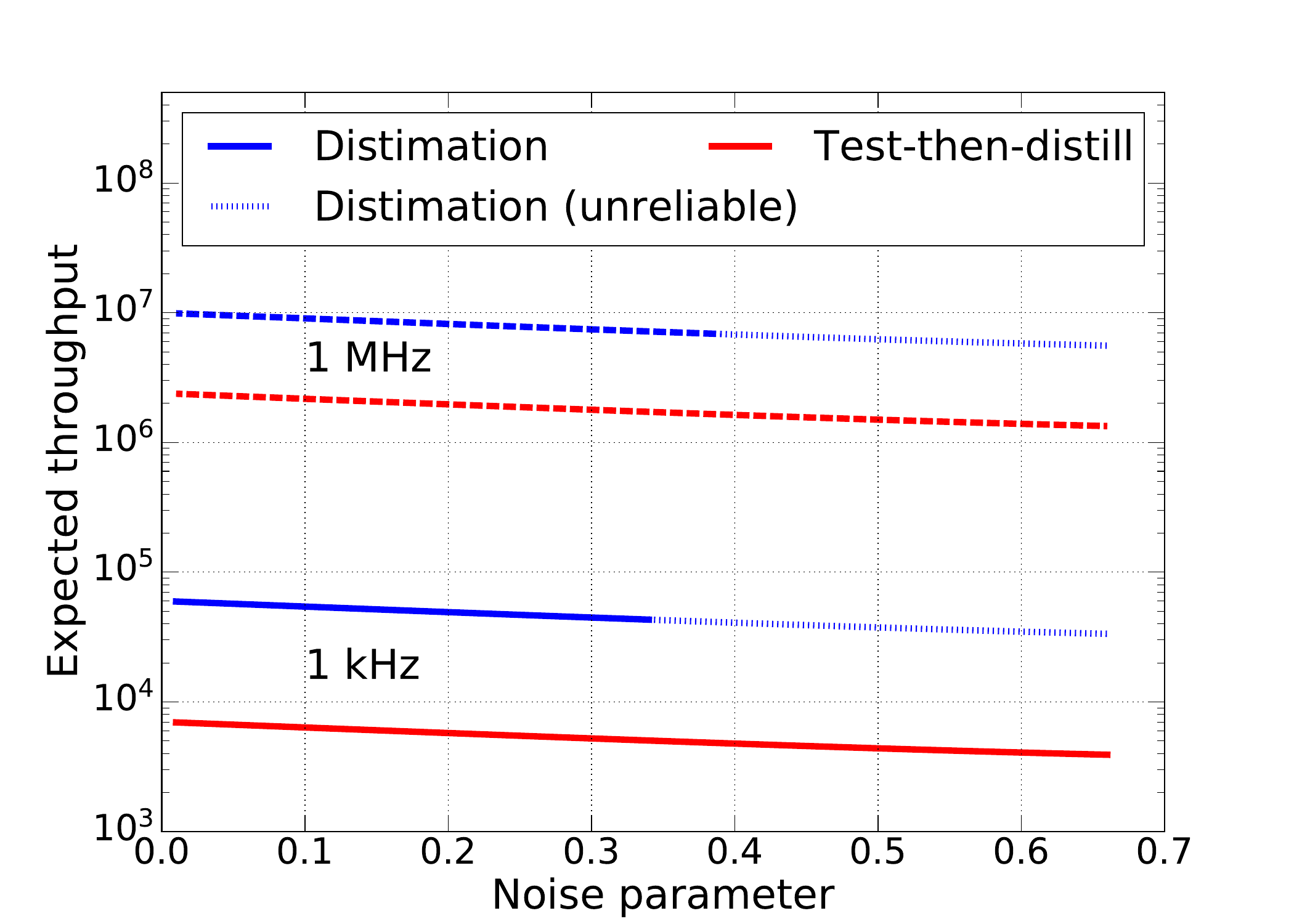}
    \caption{
    The expected throughput of distilled states using one round of the BBPSSW protocol with input isotropic states.
    Here, we fix the estimation's accuracy and failure probability. 
    We set both strategies to be active for $120$ and $20$ seconds, given the generation rates of $1$ kHz (solid red/blue lines) and $1$ MHz (dashed red/blue lines), respectively.
    In both the kilohertz (estimation error and failure probability of estimation equal to $10^{-2}$) and megahertz (estimation error and failure probability of estimation equal to $10^{-3}$) cases, there is a quantitative advantage in expected throughput when the distimation approach is used.
    However, the estimate obtained from distimation can be unreliable for high-noise parameters, since the number of available Bell pairs is insufficient to meet the target accuracy and failure probabilities.
    }
    \label{fig:performance}
\end{figure}

    Our primary metric is the expected throughput of distilled Bell pairs, as a function of the noise parameter, for fixed accuracy and failure probability. 
    We consider two different example scenarios (Figure~\ref{fig:performance}). 
    For near-term devices that can generate Bell pairs at 1 kHz, we set both strategies to be active for 120 seconds and require estimates to have an accuracy of $10^{-2}$ with a failure probability smaller than $10^{-2}$. 
    For future devices that can generate Bell pairs at 1 MHz, we set the duration to 20 seconds and require the estimates to have a stricter accuracy of $10^{-3}$ and a failure probability smaller than $10^{-3}$.

    At kilohertz generation rates, the expected distilled‑pair throughput of distimation can exceed that of the ``test‑then‑distill'' pipeline by an order of magnitude. 
    This is because every raw pair either survives as a distilled output or contributes informative syndrome data, while the baseline dedicates many Bell pairs to tomography before accruing any distillation yield.

    At megahertz generation rates, both strategies achieve the previous accuracy goals with minimal sampling overhead.
    However, as accuracy requirements tighten with advancing technology, distimation maintains a considerable throughput advantage.

\section{Challenges and open problems}

    Adopting our protocol as a routine link-layer service requires overcoming several technical hurdles, which we categorize into five broad categories.

\subsection{Impact of imperfect hardware} 

    Distimation can tolerate hardware noise by integrating noise-aware estimators, at the cost of increased sampling. 
    Currently, however, it relies on case-by-case analytical modeling. 
    For instance, previous results \cite{casapao2024distimator} demonstrated the feasibility of distimation for two-qubit distillation under independent Pauli noise and decoherence.
    Yet, a similar analysis remains absent for other existing distimation schemes \cite{casapao2025distimator,Oka2025}. 

    When local characterization is incomplete---such as omitting gate noise or Pauli decoherence effects---the estimation step may fail to converge to the correct state. 
    Recent simulation studies and small-scale demonstrations confirm this practical limitation \cite{yokomori2025evaluation}.

    A pressing challenge is therefore designing distimation techniques that remain robust when hardware models are coarse, partial, or quickly outdated, thereby eliminating the need for continual device-specific recalibration.

\subsection{{General distimation recipe}} 

    A general distimation methodology must balance three factors: (i) mapping a sufficient set of syndromes to target state parameters, (ii) managing circuit complexity, and (iii) providing provably efficient algorithms with explicit error bounds.
    
    For (i) and (ii), existing distimation strategies enrich the available syndrome data by running several complementary low-depth distillation circuits and combining their outcomes \cite{casapao2024distimator}, or by increasing the circuit width to accommodate more measurement settings \cite{casapao2025distimator,Oka2025}.
    Here, circuit complexity creates a trade-off: larger circuits characterize more states but increase qubit overhead and post-processing demands.
    Alternatively, link-layer scheduling algorithms can enrich syndrome data by extending collection into the temporal dimension.

    Regarding (iii), while the Pauli-channel learning framework by Wagner \textit{et al.}~\cite{wagner2022pauli} offers a useful foundation, adapting it to the richer landscape of distimation protocols remains an open problem. 

\subsection{Near-term deployment challenges} 

    Distimation operates under the same hardware constraints that limit near-term entanglement generation and distillation. 
    Today's quantum network nodes can only implement a limited repertoire of low-depth, low-width distillation circuits. 
    
    Complex distillation protocols requiring multiple pairs remain experimentally inaccessible. 
    In this context, protocols offering strong distimation guarantees are unsuitable if they exceed hardware capabilities or disrupt link scheduling.
    This limitation forces a trade-off between estimation performance and implementation overhead. 
    As hardware matures, however, the set of feasible distimation protocols will expand. 
    In the interim, deployments must align with current node capabilities and tailor their estimation goals accordingly.

\subsection{Time-varying quantum sources}
\label{subsec:bayesian}

    Existing distimation schemes assume source statistics remain stationary during diagnosis.
    However, hardware parameters drift over time.
    Without dynamic tracking, inferred parameters become stale, causing the control plane to make incorrect entanglement management decisions.

    For distimation to be practically useful, its estimation phase must become adaptive. 
    One option is a sliding-window estimator that constantly refreshes the noise model with the most recent syndromes while discarding older data.
    Another is to run a full-fidelity distimation sweep at start-up, followed by brief refresh bursts at regular intervals.  
    In both cases, prior estimates can serve as Bayesian priors for next updates, allowing subsequent distimation rounds to converge more quickly and consume fewer raw pairs than the initial calibration.

\subsection{Distimation in Graph–State Networks}
    
    The discussion so far has assumed a stack whose elementary resource is a Bell pair shared between neighboring nodes. 
    Alternative network proposals treat the on-demand distribution of multipartite \textit{graph states} as the primitive \cite{pirker2019quantum}, where the underlying entanglement reflects the desired multi-node connectivity.
    In such architectures, the link layer delivers small graph states directly, while higher layers fuse them into ever-larger graphs. 
    Transitioning from bipartite links to multipartite blocks expands the state space from three independent Bell-diagonal parameters to an exponential number of variables, thus reshaping the distimation problem. 
    Given these scaling constraints, distimation may be practically limited to estimating fidelity or dominant basis entries.

    Distimation depends on how the network prepares these graph states. 
    If prepared by fusing and distilling bipartite states, then distimation can remain link-local.
    The fusion stage would include a fidelity bookkeeping scheme that compiles link-level estimates into a global quality metric.
    Alternatively, if the network utilizes direct multipartite distillation steps, it requires novel distimation routines adapted to these operations.

\section{Summary}

    Distimation, the unification of channel-state estimation and entanglement distillation within a single link-layer protocol, offers a compelling path toward resource-efficient quantum networks. 
    By reusing the classical syndromes that distillation inherently generates, distimation eliminates the dedicated ``test-then-distill'' cycle that dominates today’s state-of-the-art stacks.
    Our framework simplifies the protocol stack while increasing usable entanglement throughput. 
    Building on this, we identified several open challenges: accurate local-device calibration, adaptation to time-varying sources, and extending distimation to multipartite entanglement distribution. 
    Results presented here suggest that distimation could serve as a standard building block for future quantum networks, lowering resource costs while streamlining network operations.

\bibliographystyle{IEEEtran}
\bibliography{reference.bib}


\end{document}